\documentclass[dvips,12pt]{article}
\usepackage{graphicx}
\begin{document}
\markboth{}{Motion of a Vector Particle in a Curved Space-time
III}

\title{Motion of a Vector Particle in a Curved Space-time.
III Development of Techniques of Calculations}

\author{Turakulov Z. Ya.\\
\footnotesize National University of  Uzbekistan, Vuzgorodok, Tashkent
700174 Uzbekistan\\ \footnotesize Astronomy Institute, and Institute
of Nuclear Physics, Ulughbek, Tashkent
702132 Uzbekistan\\ \footnotesize Tashkent, Uzbekistan\\
\\
Muminov A. T.
\\ \footnotesize Institute of Applied Physics affiliated to the
National University of  Uzbekistan\\ 
\footnotesize Vuzgorodok 3A, Tashkent, 700174 Uzbekistan\\
\footnotesize{amuminov2002@mail.ru}}

\maketitle

\begin{abstract}
The studies of influence of spin on a photon's motion in a
Schwartzschild spacetime is continued. In the previous paper
\cite{zr2} the first order correction to the geodesic motion is found
for the first half of the photon world line. The system of equations
for the first order correction to the geodesic motion is reduced to a
non-uniform linear ordinary differential equation. The equation
obtained is solved by the standard method of integration of the Green
function.
\end{abstract}

\def \vc {\vec}
\newcommand{\T}{\tilde }
\newcommand{\be}{\begin{equation}}
\newcommand{\ee}{\end{equation}}
\newcommand{\lr}{\left(}
\newcommand{\rr}{\right)}
\newcommand{\lfc}{\left\{}
\newcommand{\rtc}{\right\}}
\newcommand{\lta}{\left\langle}
\newcommand{\rta}{\right\rangle}
\newcommand{\too}{\longrightarrow}
\newcommand{\eps}{\varepsilon}
\newcommand{\vphi}{\varphi}
\newcommand{\fracm}[2]
{{\displaystyle#1\over\displaystyle#2\vphantom{{#2}^1}}}
\newcommand{\sprd}[2]
{\left\langle#1\,,#2\right\rangle\,}
\newcommand{\prt}{\partial}
\def \elmg {electromagnetic }
\def \va {\vc A }
\def \vac {{\vc A }^*}
\def \ca {\bar{A}}
\def \vcx {\dot{\vc x }}
\def \vdx {\delta\vc x }
\def \vca {\dot{\vc A }}
\def \vcca {\dot{{\va\vphantom{A}\,}^*}}
\def \wd {\wedge}
\newcommand{\Ds}[1]{{D#1\over ds}}
\def \dds {\frac{d}{ds}}
\newcommand{\lcom}[1]{{\left[#1\right.}}
\newcommand{\rcom}[1]{{\left.#1\right]}}
\newcommand{\lacm}[1]{{\left\{#1\right.}}
\newcommand{\racm}[1]{{\left.#1\right\}}}
\newcommand{\cc}[1]{\bar{#1}}
\newcommand{\half}{\frac{1}{2}}
\newcommand{\quart}{\frac{1}{4}}
\def \oh {{1/2}}
\def \moh {{-1/2}}
\def \goo {{1-r_g/r}}
\def \gob {(1-r_g/r)}
\def \rd {{r^2-D^2\gob}}
\def \rdi {{r_1^2-D^2(1-r_g/r_1)}}
\newcommand{\ctg}[1]{\tan^{-1}{#1}}
\def \vce {{\vc e }}
\def \vcn {{\vc n }}
\def \cdt {\hspace{0.03em}\cdot\hspace{0.03em}}
\def \fdt {\hspace{0.36em}}
\def \dtx {\dot{x}}
\def \dta {\dot{A}}
\def \dx {{\delta x}}
\newcommand{\gamm}[3]{\,\gamma_{#1#2\cdt }^{\fdt\fdt#3}\,}
\newcommand{\con}[2]{\omega_{#1\cdt}^{\fdt#2}}
\newcommand{\cur}[2]{\Omega_{#1\cdt}^{\fdt#2}}
\def \ccn {\mbox{C.C. }}
\newcommand{\dd}[2]{\fracm{\prt#1}{\prt#2}}
\newcommand{\der}[2]{\frac{d#1}{d#2}}
\def \aRL {\left|\dd{R}{L}\right|}
\newcommand{\sg}[1]{{\varepsilon(#1)}}
\newcommand{\stick}[2]{{\left.#1\right|_{#2}}}
\section{Introduction}

Existence of spin-gravitational interaction was proved by
A.~Papapetrou in his analysis of motion of deformable body
\cite{pap,cor-pap}. More recently we have derived Papapetrou equations
as a reduction of equations of motion of (tangent) rigid body
\cite{th017}. However, analysis based on classical mechanics does not
admit passage to the limit of zero mass. The problem of motion of a
massless spinning particle is especially interesting due to importance
of electromagnetic waves in astronomial observations. Importance of
the problem of photon world line with account of the
spin-gravitational interaction was under discussion for decades
\cite{bm1}-\cite{bm4}, however it could not be formulated properly
until equations of motion for massless spinning particle were derived.

Passage to classical mechanical limit from a non-scalar field
equation is an alternative approach tried by numerous authors
\cite{th503}-\cite{baylin}. Straightforward derivation of photon's
equations of motion from Lagrangian of electromagnetic field was
made in our works \cite{zr1,za}. As the result we have obtained
Papapetrou equation in the form\begin{equation}
\label{papa}\omega\Ds{\dtx^a}=R_{\alpha\beta\cdt c}^{\fdt\,\fdt
a}\, \dot x^c\sigma^{\alpha\beta},\end{equation} where $\omega$ is
frequency and $\sigma^{\alpha\beta}$ -- spin of photon which lies
on tangent vector subspace of polarization vectors $\vcn_\alpha,$
$\alpha,\beta,\ldots=2,3.$ The spin remains parallel to itself in
the polarization subspace along the world line. It is seen from
this equation that the effect of spin-gravitational interaction
grows linearly with the wavelength, so, under large enough
wavelengths it may well be observable.

The next step is to find out photon world lines in a given model of
space-time. The simplest way is to construct world lines via computing
the first order correction to isotropic geodesic. These computation in
case of Schwartzschild space-time was completed in our recent work
\cite{zr2} where, however, the desired result was not reached because
the method used is valid only on a half of the world line which starts
from infinitely distant source and ends up at the minimal value of the
radial coordinate $r$ (periastr). To construct the whole world line
one needs to match two such halves built in different coordinate
systems that is quite difficult. The goal of the present work is to
work out a method of building the whole of the world line of a
massless spin 1 particle in Schwartzschild space-time.

In the framework of the accepted approximation scheme we chose an
isotropic geodesic (hereafter: reference geodesic) which is a world
line of a spinless particle, then include spin into the right-hand
side of the equation~(\ref{papa}) putting it parallel to itself along
the geodesic. It is convenient to introduce a geodetic flow in the
equatorial plane ($\theta=\pi/2$ in spherical coordinates) as a vector
field of velocities and its small variation. By construction the
vector field satisfies geodesic equation, so it remains to find out
its small variation such that the total vector satisfies the equation
(\ref{papa}). This yields a linear equation  of second order for the
variation which can be solved on a geodesic of the flow. Solution of
this equation is the first order correction to be found. The only
independent variable in the equation is parameter $s$ on the reference
geodesic.

The metric of the Schwarzschild space-time can be specified by a field
of orthonormal frame. A sample of pair of dual to each other covector
and vector standard (with no isotropic elements) orthonormal frames is
\begin{equation}\label{onf}\left.\begin{array}{rcl}
\theta^0&=&(1-r_g/r)^\oh\,dt,\\ \theta^1&=&(1-r_g/r)^\moh\,dr,\\
\theta^2&=&rd\theta,\\ \theta^3&=&r\sin{\theta}\,d\vphi;\\
\end{array}\right.\quad\left.\begin{array}{rcl}
\vce_0&=&(1-r_g/r)^\moh\,\prt_t,\\ \vce_1&=&(1-r_g/r)^\oh\,\prt_r,\\
\vce_2&=&r^{-1}\prt_\theta,\\ \vce_3&=&(r\sin{\theta})^{-1}\,
\prt_\vphi\\ \end{array}\right.\end{equation} where orthonormality of
the frames means $$<\theta^a,\theta^b>=\eta^{ab},
\quad<\vce_a,\vce_b>=\eta_{ab},\quad\theta^a(\vce_b)=\delta^a_b,$$
$\eta_{ab}$ the standard Minkowski metric. The corresponding
connection 1-form is\begin{equation}\label{con}
\left.\begin{array}{rcl}\vspace{0.18cm}
\con{1}{0}&=&\fracm{r_g}{2r^2\sqrt{1-r_g/r}}\,\theta^0,\\
\vspace{0.15cm}\con{2}{0}&\equiv&0,\\\con{3}{0}&\equiv&0;\\
\end{array}\right.\quad\left.\begin{array}{rcl}\vspace{0.15cm}
\con{2}{1}&=&-\fracm{\sqrt\goo}{r}\,\theta^2,\\ \vspace{0.15cm}
\con{3}{2}&=&-\fracm{\ctg{\theta}}{r}\,\theta^3,\\
\con{1}{3}&=&\fracm{\sqrt\goo}{r}\,\theta^3.\\ \end{array}\right.
\end{equation} The 2-form of curvature referred to the frame is:
\begin{equation}\label{cur}\left.\begin{array}{rcl}\vspace{0.15cm}
\cur{1}{0}&=&\fracm{r_g}{r^3}\,\theta^0\wd\theta^1,\\ \vspace{0.15cm}
\cur{2}{0}&=&-\fracm{r_g}{2r^3}\,\theta^0\wd\theta^2,\\
\cur{3}{0}&=&-\fracm{r_g}{2r^3}\,\theta^0\wd\theta^3;\\
\end{array}\right.\quad\left.\begin{array}{rcl}\vspace{0.15cm}
\cur{2}{1}&=&-\fracm{r_g}{2r^3}\,\theta^1\wd\theta^2,\\
\vspace{0.15cm}\cur{3}{2}&=&\fracm{r_g}{r^3}\,\theta^2\wd\theta^3,\\
\cur{1}{3}&=&-\fracm{r_g}{2r^3}\,\theta^3\wd\theta^1.\\
\end{array}\right.\end{equation}\section{Isotropic geodesic flow,
canonical parameter and the co-moving frame}

We construct a congruence of isotropic geodesic lines lying wholly on
$\theta=\pi/2$ hypersurface (an equatorial plane) by solving the
Hamilton-Jacobi equation for isotropic geodesics on it
\begin{equation}\sprd{d\Psi}{d\Psi}=0\label{HJ}\end{equation}where the
function to be found is of the form \begin{equation}\label{psi}
\Psi=Et-L\vphi+R(r)\end{equation}and contains two constant parameters
$E$ and $L$. Accordingly to Hamilton-Jacobi theorem the integrated
equations of geodesics are $$\frac{\prt\Psi}{\prt E}=const,\quad
\frac{\prt\Psi}{\prt L}=const.$$These equations specify the variables
$t$ and $\vphi$ as functions of the variable $r$:\begin{equation}
\label{eq4geo}t=-\dd{R}{E},\quad\vphi=\vphi_0+\dd{R}{L},\quad
\theta\equiv\pi/2\end{equation}where the function $R$ is found by
integrating the Hamilton-Jacobi equation~(\ref{HJ}):\begin{equation}
\label{R} R(r)=-\varepsilon(t)\int_{r_0}^r
\fracm{\sqrt{E^2r^2-L^2(\goo)}}{r(\goo)}dr,
\end{equation}where $\varepsilon(t)$ is the well-known step function
equal +1 for positive $t$ and -1 for negative $t$.

Each geodesic of the congruence starts at $r=\infty$ under $t=-\infty$
and reaches the periastr $r=r_0$ at the moment $t=0$. After this
moment the cordinate $r$ grows and reaches infinity under $t\to+
\infty$. We choose initial conditions such a way that
$\stick{\vphi}{t\to-\infty}$ $=0$, so that $\vphi_0=$
$\aRL_{r\to\infty}$. It must be noted that the isotropic geodesics of
the congruence lie wholly on surfaces of level of the function $\Psi$
or, in other words, $\Psi$ is constant on each of them. Therefore this
function cannot be used as a parameter on the geodesics. Instead, a
canonic parameter on the geodesics will be introduced below.

In order to reduce the number of non-zero components of spin of
the particle it is convenient to introduce a co-moving frame on
the geodesics like that done in work \cite{zr2}. Since the
geodesics are isotropic the co-moving frame contains isotropic
elements $\vcn_\pm$. One of them, $\vcn_-$ is orthogonal to
surfaces $\Psi=const$, (thus, due to properties of isotropic
vectors, tangent to them everywhere):$$\vcn_-=
\sprd{dx^i}{d\Psi}\prt_i$$ and another isotropic vector $\vcn_+$
is chosen only due to the normalization condition
$\sprd{\vcn_-}{\vcn_+}$ $=1$. Two other vectors which constitute
polarization subspace $\vcn_\alpha,$ $\alpha=2,3$ are normalized
to unit, space-like, orthogonal to each other and to the isotropic
vectors:\begin{equation}
\begin{array}{rcl}\label{comov}\vspace{0.15cm}
\vcn_-&=&\gob^{-1}\,\prt_t+Dr^{-2}\,\prt_\vphi
-R'E^{-1}\gob\,\prt_r,\\ \vspace{0.15cm}
\vcn_+&=&\gob/2\,\left[\gob^{-1}\,\prt_t-Dr^{-2}\,\prt_\vphi
+R'E^{-1}\gob\,\prt_r\right],\\
\vcn_2&=&r^{-1}\,\prt_\theta,\\
\vcn_3&=&(\rd)^oh\left(r^{-2}\,\prt_\vphi-\dd{R'}{L}\gob\,
\prt_r\right),\\ \end{array}\end{equation}where$$R'=\varepsilon(t)
\fracm{\sqrt{E^2r^2-L^2(\goo)}}{r(\goo)}$$  and $D=L/E$ is the
impact parameter. The vector $\vcn_-$ is obtained from the 1-form
of momentum $d\Psi$, by index lifting therefore, it satisfies the
geodesic equation:$$
\vcn_-\circ\vcn_-\equiv(\vcn_-\cdot\nabla)\vcn_-\equiv0$$
and hence $\vcn_-$ is the canonical velocity on the geodesic
\cite{0geo}. Thus, the canonical parameter can be found by
equating $\vcn_-$ to $\dtx=$ $dx^i/ds\,\prt_i$ that gives:
$$\der{r}{s}=-R'/E\,\gob,\quad\frac{d}{ds}=\der{r}{s}\der{}{r}.$$
It is seen that sign of $dr/ds$ changes when passing through the
hypersurface $t=0$ which separates the isotropic geodesic into two
halves. Under $t<0$ coordinate $r$ decreases along the geodesic
until at $t=0$ it reaches its minimal value $r_0$ and afterwards grows
up to infinity under $t>0$. Therefore $r$ can be used as a parameter
only on a half of the geodesic where the coordinate $r$ runs the range
$r\in(r_0,\infty)$.

In order to compute the first order correction to the whole of the
geodesic we need a monotonic parameter $s$ on it. It is convenient
to define it such that at the moment of time $t=0$ when the
coordinate $r$ takes its minimal value ($r=r_0$) the value of $s$
is zero, for example: \be s=\varepsilon(t)\int\limits_{r_0}^r
\fracm{r\,dr}{\sqrt{\rd}}.\label{sofr}\ee Now we can find the relation
between two parameterizations on the geodesic for each of its halves:
\begin{equation}\label{s2r}ds=\fracm{\varepsilon(s)r\,dr}{\sqrt{\rd}}.
\end{equation} Since hereafter we shall use the values of the function
$R$ on the geodesic it is convenient to introduce an alternating
function $R(s)$ with its derivative w.r.t $L$ :\begin{equation}
\label{RL} \dd{R(s)}{L}=\varepsilon(s)\left|\dd{R(r)}{L}\right|,
\hspace{1cm}\aRL=\int_{r_0}^r\fracm{D\,dr}{r\sqrt{\rd}}.\end{equation}

Besides, we need expressions of vectors $\vcn_p=$ $n_p^a\vce_a$ via
vectors $\vce_a$ of standard frame:\begin{equation}
\begin{array}{rcl}\label{comov2}\vspace{0.15cm}
\vcn_-&=&\gob^\moh\,\vce_0+Dr^{-1}\,\vce_3
-R'E^{-1}\gob^\oh\,\vce_1,\\\vspace{0.15cm}
\vcn_+&=&\gob/2\,\left[\gob^\moh\,\vce_0 Dr^{-1}\,\vce_3
+R'E^{-1}\gob^\oh\,\vce_1\right],\\
\vcn_2&=&\vce_2,\\
\vcn_3&=&(\rd)^\oh\left(r^{-1}\,\vce_3-\dd{R'}{L}\gob^\oh\,\vce_1
\right).\\ \end{array}\end{equation}\section{First order approximation
of the Papapetrou equation}

Assuming that interaction of spin with gravitation is weak and
deviation of the ray from the reference geodesic: $\dx^a/r\ll1$ is
small we shall derive the equation for the deviation from the
Papapetrou equation and solve it. It must be noted that analysis
made in the work \cite{zr2} contains one more inaccuracy: when
considering deviation from a geodesic one should take into account
the difference between values of connection on the geodesic and on
the deviation from it. If it is done properly instead of the
l.h.s. of the geodesic equation $\vcn_-\circ\vcn_-$ one has that
of the Jacobi equation$$ \frac{D^2}{ds^2}\dx^a-R_{b\cdt cd}^{\fdt
a}\,n_-^b\,n_-^c\,\dx^d$$ \cite{ko-no}. Therefore the desired
equation for the small deviation $\dx^a$ extracted form the
Papapetrou equation~(\ref{papa}) is
\begin{equation}\label{papa1gen}
\left[\frac{D^2}{ds^2}\dx^a-R_{b\cdt cd}^{\fdt
a}\,n_-^b\,n_-^c\,\dx^d\right]=-\lambda R_{b\cdt cd}^{\fdt
a}\,n_-^b\,n_2^c\,n_3^d\,\sigma^{23},
\end{equation}where $\lambda=\omega^{-1}=const$ is the (asymptotical)
wavelength.

It was shown in the work \cite{zr2} that the world line of a photon
lie initially in the equatorial plane $\theta=\pi/2$ the vector of
deviation has in the frame~(\ref{comov2}) the only non-zero component
$\dx^2$. Therefore it is convenient to have explicit form of the
following contractions of the Riemann tensor:\begin{eqnarray}
\label{contract}
R_{b\cdt c2}^{\fdt2}\,n_-^b\,n_-^c=-\frac{3r_gD^2}{2r^5};\\
R_{b\cdt cd}^{\fdt2}\,n^b_-\,n_2^c\,n_3^d
=\frac{3r_gD}{2r^5}\sqrt{\rd}.\end{eqnarray} The latter is needed
because in the co-moving frame~(\ref{comov2}) the spin of photon
has only one non-zero component $\sigma^{23}$. Taking into account
that in the co-moving frame all components of the connection
1-form are zero we substitute partial derivative of the deviation
vector for its covariant derivative and, referring to~(\ref{cur})
and~(\ref{comov2}) we reduce the Papapetrou equation for the
deviation vector in Schwarzschild background to the following
ordinary differential
equation$$\frac{d^2\,y}{ds^2}+\frac{3r_gD^2}{2r^5}\,y=
-\frac{3\lambda r_gD}{2r^5}\sqrt{\rd}\sigma^{23},$$where the
function $y(s)$ to be found stands for $\dx^2$. Accordingly to
fact that under $s\to-\infty$ the ray coincides with reference
geodesic sought solution of the eq. must obey asymptotical conditions
as follows:\begin{equation}\label{ascon}
y(-\infty)=0,\quad \frac{d\,y}{ds}(-\infty)=0.
\end{equation}

It is convenient to rewrite the approximated Papapetrou equation
in the standard denotions\begin{equation}\label{jac-nu}
\frac{d^2y}{ds^2}+Q(s)y(s)=F(s),\end{equation}with
\begin{eqnarray}Q(s)&=&\fracm{3r_gD^2}{2r^5},\nonumber\\
F(s)&=&-C\,\fracm{D\sqrt{\rd}}{r^5},\\
C&=&3\lambda r_g\sigma^{23}/2\nonumber.
\end{eqnarray}This is a well-known linear non-uniform ordinary
differential equation of second order which can be solved by standard
methods. To do it we need first the solution of the uniform equation
which corresponds to commonplace deviation of geodesics, thus, to the
Jacobi equation in which spin-gravitational is `switched off' by
putting $C=0$. Then as the uniform equation is solved we can use the
solution for constructing solution of the non-uniform equation and
obtain deviation caused by the spin-gravitational interaction.
\section{Integration of the Jacobi equation}

In this section we make straightforward calculation of the deviation
vector which in the field of local frames~(\ref{comov2}) has only one
non-zero component $\dx^2$. For this end we consider a geodesic flow
with small 2-component found from the corresponding solution of the
Hamilton-Jacobi equation. Unlike the solutions considered above this
solution depends on all four coosdinates\begin{equation}\label{tPsi}
\T\Psi=Et-M\vphi+R(r)+\Theta(\theta)\end{equation} and contains three
arbitrary constants $E,\, L$ and $M$. The function $R(r)$ is the same
as before and the new function $\Theta(\theta)$ has the form $$
\Theta(\theta)=\pm\int^\theta\sqrt{L^2-M^2/\sin^2\theta}.$$ If we put
$L=M$ theis function vanishes and the corresponding geodetic flow
reduces to the flow considered above. So, if we put
$(L^2-M^2)/L^2=\eps^2$ an infinitesimal parameter we obtain a family
of geodetic flows which are infinitesimally close to the geodetic flow
in the equatorial plane. Applying the Hamilton-Jacobi theorem to the
solution (\ref{tPsi}) gives:
\begin{eqnarray}\label{e4ng1}\dd{\T\Psi}{M}=-\vphi\mp\int^\theta
\frac{M/sin^2\theta}{\sqrt{L^2-M^2/\sin^2\theta}}\,d\theta=const,\\
\nonumber\dd{\T\Psi}{L}=\pm\int^\theta
\frac{L\,d\theta}{\sqrt{L^2-M^2/\sin^2\theta}}+\dd{R}{L}=const
\end{eqnarray} and all the rest equations remain unchanged.

The second line specifies the component $\delta\theta$ of the
deviation$$\delta\theta=\frac{\sqrt{L^2-M^2}}{L}\sin\left(\dd{R}{L}+
const\right)=\eps\sin\left(\dd{R}{L}+const\right)$$ and the first line
contains the same integral, therefore, neglecting the second poweer of
the parameter $\eps$ we obtain the same equation as under $\eps=0$:
$$-\vphi+M/L\,\dd{R}{L}=const,$$consequently, the component
$\delta\vphi$ of the deviation is zero. Thus, the deviation vector has
only one non-zero component which is $\delta\theta$:
$$\delta r=\delta t=\delta\vphi\equiv0,\quad\delta\theta=
\eps\sin\left(\dd{R}{L}+\zeta\right),$$ where $\zeta=const$ and $\eps$
is an abitrary infinitesimal parameter. Parameter $\zeta$ specifies
deviation from the geodesic at $t=-\infty$ which can also be put
non-zero. In our denotions $y=\delta x^2$ solution of the Jacobi
equation just obtained is:\begin{equation}\label{sol-Jac}
y=\eps r\sin\left(R_L+\zeta\right)\end{equation}where
$\left|R_L(r_1)\right|$ stands for the elliptical integral (\ref{RL})
\section{Integration of the non-uniform equation}

In this section we solve the non-uniform equation~(\ref{jac-nu}) by 
standard method of integrating the Green function. By definition the
Green function of this ordinary differential equation (\ref{jac-nu})
is a piecewise-smooth function composed of two solutions of the
uniform equation. Since the differential equation is of second order
the relevant Green function $G(s|s_1)$ should be continuous in the
point $s=s_1$ wile its first derivative makes a unit jump \cite{MF}.
This way we construct the desired Green function by matching two
solutions of the uniform equation as follows:\begin{equation}
\label{green}G(s|s_1)=\frac{r_1}{D}\left\{ \begin{array}{l}
0\quad\mbox{if}\quad s<s_1,\\ r\sin\left\{\sg{s}\left|R_L(r)\right|-
\sg{s_1}\left|R_L(r_1)\right|\,\right\} \quad\mbox{if}\quad s_1<s;\\
\end{array}\right.\end{equation}Here the factor $r_1/D$ normalizes the
jump of the derivative, $r_1=r(s_1)$, $r=r(s)$ are specified by the
function inverse to the function $s(r)$ defined in the equation
(\ref{sofr}) and the variable $s$ enters only the step function
$\varepsilon(s)$ which specifies only the sign depending on the half
of the world line. Besides, solution under $s<s_1$ is trivial that
provides validity of our asymptotical conditions (\ref{ascon}) at
$s\to-\infty$.

Now, following the standard procedure we can find out solution of the
nonuniform Jacobi equation (\ref{jac-nu}) as follows: \be
\begin{array}{l}\label{y1}\displaystyle
y(s)=\int\limits_{-\infty}^\infty G(s|s_1)F(s_1)ds_1=\\
\displaystyle=-Cr\!\int_{-\infty}^s\!
\fracm{\sin\left\{\sg{s}\left|R_L(r)\right|-
\sg{s_1}\left|R_L(r_1)\right|\,\right\}}{r_1^4}\sqrt{\rdi}\,ds_1.
\end{array}\end{equation}
Our goal is to describe behavior of the solution obtained under $s>0$
corresponding to deviation the ray off the reference geodesic after
the periastr. It is convenient to represent it by the angle of
deviation: \begin{eqnarray*}\delta\theta(s)=\fracm{y(s)}{r}=
-C\!\int_{-\infty}^{s}\!
\fracm{\sin\left\{\varepsilon(s)\left|R_L(r)\right|-
\varepsilon(s_1)\left|R_L(r_1)\right|\,\right\}}{r_1^4}\sqrt{\rdi}\,
ds_1=\\\int\limits_{-\infty}^0\hspace{-0.4em}(\ldots)ds_1+
\!\int\limits_{0}^s\!(\ldots)ds_1\end{eqnarray*} where we divide the
domain of integration into parts of definite signs of the step
function of $s_1$:$$\delta\theta(s)=
\int_{\infty}^{r_0}(\ldots)\fracm{-r_1\,dr_1}{\sqrt{\rdi}}+
\int_{r_0}^r(\ldots)\fracm{r_1\,dr_1}{\sqrt{\rdi}}.$$Here the square
roots in the denominators appear when passing from the varibale $s_1$
to $r_1$ (\ref{s2r}. At the same time we substitute the limit
$=\infty$ of $r_1$ for the limit $-\infty$ of $s_1$. After cancelling
the square roots the expression simplifies and takes the form$$
\delta\theta(s)=-C\int_{r_0}^\infty\fracm{\sin\left\{\left|R_L(r)\right|+
\left|R_L(r_1)\right|\,\right\}}{r_1^3}\,dr_1
-C\int_{r_0}^r\fracm{\sin\left\{\left|R_L(r)\right|-
\left|R_L(r_1)\right|\,\right\}}{r_1^3}\,dr_1\,.$$ 
Introducing the notion of asymptotical angle $\delta\theta_\infty=$
$\delta\theta(+\infty)$ which describes the final angular divergence
between the ray and its reference geodesic we obtain expression for
the deviation on $s>0$ branche of the ray as follows:
$$y(s)=r\delta\theta_\infty
+r\frac{\sin|R_L(r)|-\sin|R_L(\infty)|}{\sin|R_L(\infty)|}
\delta\theta_\infty+Cr
\int_r^\infty\fracm{\sin\left\{\left|R_L(r)\right|-
\left|R_L(r_1)\right|\,\right\}}{r_1^3}\,dr_1.$$
Evidently, the last term vanishes under $s\to\infty$. Now let's find
expression for $\delta\theta_\infty$. After some algebraic operations
we obtain:\be\label{asang}
\stick{\delta\theta}{s\to\infty}=-2C\sin\left|R_L(\infty)\right|
\int_{r_0}^\infty\cos\left|R_L(r_1)\right|\hspace{0.5em}
r_1^{-3}\,dr_1.\end{equation}

It is seen that the integral $\left|R_L(r_1)\right|$ cannot be
expressed in terms of convenient functions. However, since we are
seeking for the first order on $r_g/r_0\ll1$ and the constant factor
$C=$ $3r_g\sigma^{23}/2E$ already contains the factor $r_g$, it
suffices to take the integral in zeroth order approximation on
$r_g/r_0$. In other words, to obtain the desired result we can put
$r_g=0$ in $\left|R_L(r_1)\right|$. This is possible because though
the integrand has square root singularity at $r_1=r_0$ all the
integrals in (\ref{asang}) are convergent. Moreover, the integrals as
functions of $r_g$ are continuous and differentiable at $r_g=0$.

Now, simplifying the integral (\ref{RL}) we obtain$$
{\left|R_L(r)\right|}_{r_g=0}=\arccos(r_0/r).$$ In this approximation
we have $\left|R_L(\infty)\right|=\pi/2.$ Substituting these
expressions into (\ref{asang}) simplifies the integration and finally
we have:$$\stick{\delta\theta}{s\to\infty}=-2C\int_{r_0}^\infty
\cos[\arccos(r_0/r_1)]\fracm{dr_1}{r_1^3}=
-2C\int_{r_0}^\infty\frac{r_0\,dr_1}{r_1^4}=
-\frac{r_g\sigma^{23}}{Er_0^2}=-\frac{r_g\sigma^{23}}{ED^2},$$
where we take into account that $\stick{r_0}{r_g=0}=D.$
Substituting this estimation to formula for the deviation
we obtain under $s>0$ following asymptotical presentation:
$$y(s)=r\delta\theta_\infty+
\frac{D^2}{2r}\delta\theta_\infty+O(Cr^{-2}).$$This expression is
valid for the second half of the world line and represents photon
world line under large $r$. Since the deviation on the second half
grows linearly even if there is no interaction, just because the
initial plane of the ray is lost during the first half, this linear
growth yields about one order grater deviation at infinity than that
on the periastr obtained in the work \cite{zr2}.
\end{document}